\documentclass[10pt,conference]{IEEEtran}
\usepackage{amsmath,amssymb}
\usepackage{latexsym}
\usepackage{comment}
\usepackage{graphicx}
\usepackage{subfigure}

\newcommand{\Case}[1]{\noindent{\em Case} #1: }

\newtheorem{lemma}{\bf Lemma}
\newtheorem{theorem}[lemma]{\bf Theorem}
\newtheorem{corollary}[lemma]{\bf Corollary}

\ifCLASSINFOpdf
\else
\fi
\begin{document}
%
\title{Energy Efficient Geographical Load Balancing via Dynamic Deferral of Workload}

\author{\IEEEauthorblockN{Muhammad Abdullah Adnan, Ryo Sugihara\IEEEauthorrefmark{1} and Rajesh Gupta}
\IEEEauthorblockA{University of California San Diego, CA, USA; \quad \IEEEauthorrefmark{1}Amazon.com\\
\{madnan,rgupta\}@ucsd.edu; \quad \IEEEauthorrefmark{1}sugiryo3@gmail.com
}}

\maketitle

\begin{abstract}
With the increasing popularity of Cloud computing and Mobile computing, individuals, enterprises and research centers have started outsourcing their IT and computational needs to on-demand cloud services. Recently geographical load balancing techniques have been suggested for data centers hosting cloud computation in order to reduce energy cost by exploiting the electricity price differences across regions. However, these algorithms do not draw distinction among diverse requirements for responsiveness across various workloads. In this paper, we use the flexibility from the Service Level Agreements (SLAs) to differentiate among workloads under bounded latency requirements and propose a novel approach for cost savings for geographical load balancing. We investigate how much workload to be executed in each data center and how much workload to be delayed and migrated to other data centers for energy saving while meeting deadlines. We present an offline formulation for geographical load balancing problem with dynamic deferral and give online algorithms to determine the assignment of workload to the data centers and the migration of workload between data centers in order to adapt with dynamic electricity price changes. We compare our algorithms with the greedy approach and show that significant cost savings can be achieved by migration of workload and dynamic deferral with future electricity price prediction. We validate our algorithms on MapReduce traces and show that geographic load balancing with dynamic deferral can provide 20-30\% cost-savings.
\end{abstract}




%
\IEEEpeerreviewmaketitle

\section{Introduction}
The increase in energy prices along with the rise of cloud computing, brings up the issue for making clouds energy efficient; as according to an EPA report, servers and data centers consumed 61 billion Kilowatt at a cost of \$4.5 billion \cite{EPA_Report}. Moreover, the ability to dynamically track electricity price variations due to enhancements to the electrical grid, raise the possibility of utilizing cloud computing for energy efficient computing. Recently there has been a lot of exploration on this topic, searching for opportunities to reduce energy consumption in the context of cloud \cite{beloglazov_survey, gandhi, GLB_Wierman,Qureshi,rao_market,Buchbinder}. While there are a number of hardware and software techniques for energy savings considering different aspects, one non-conventional perspective is to utilize the predetermined service level agreements (SLAs) for energy efficiency. Often there is flexibility in the specification of SLAs and the system could use that flexibility to improve the performance and efficiency \cite{Deadline_Ballani, garg_sla}. Specifically, latency is an important performance metric for any web-based services and is of great interest to service providers who are responsible for services on the cloud. The goal of this paper is to utilize the delay or latency requirements to make cloud computing more energy efficient.

Naturally, energy efficiency in the cloud has been pursued in various ways including the use of renewable energy \cite{stewart_renewable, GLB_renewable,zhang2,Le} and improved scheduling algorithms \cite{GLB_Wierman,Buchbinder,Dynamic_RS,pedram}, etc. Among them, improved scheduling algorithm is a promising approach for its broad applicability regardless of hardware configurations. The idea of utilizing SLA information to improve performance and efficiency is not entirely new. Recent work explores utilization of application deadline information for improving the performance of the applications (e.g. see \cite{Deadline_Ballani, garg_sla}). But the opportunities for energy efficiency remain unexplored. In this paper, we utilize the flexibility from the Service Level Agreements (SLAs) for different types of workload to reduce energy consumption.

\begin{figure}
\begin{center}
\includegraphics[width=3.4in]{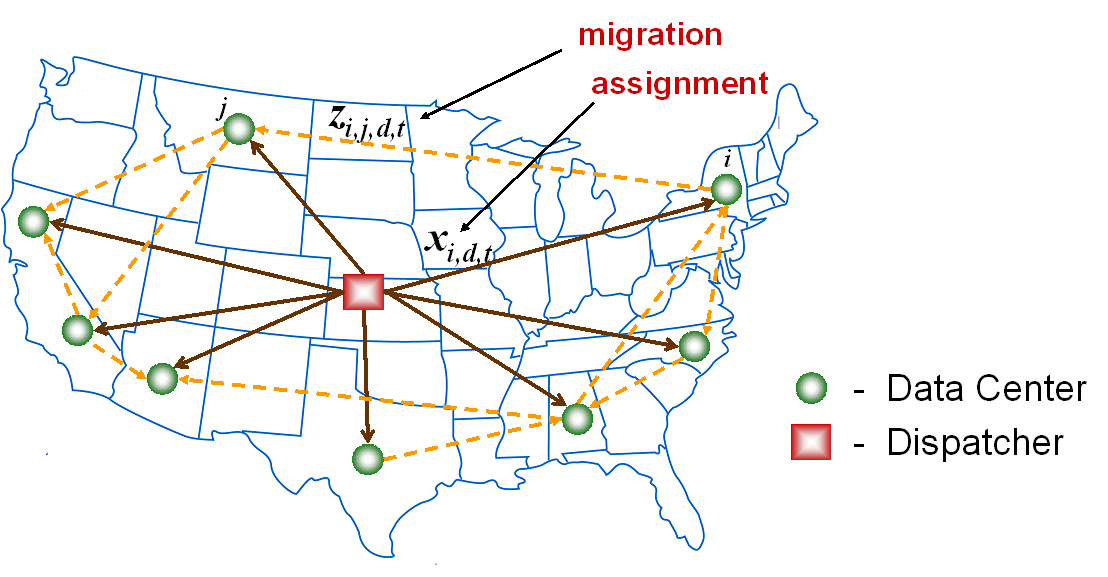}
\caption{Geographical Load Balancing.}
\label{fig:glb}
\end{center}
\end{figure}

We consider the problem of geographical load balancing in the cloud (Figure~\ref{fig:glb}). In cloud computing, each center of execution (data centers) are usually located in different geographic locations which are often in different time zones. Due to the increase in cost of energy, the electric billing companies have different pricing rates for electricity at different locations and at different times of the day. Hence load balancing decisions should take into account the current time zones and locations of data centers during task assignment for minimizing the total cost of energy consumption in the cloud. In this paper, we investigate and analyze how the pricing of energy at different times of the day along with `task migration' account for the decisions for task assignment and deferral in the cloud. We use deadline information to defer some tasks so that we can reduce the total cost for energy consumption for executing the workload depending on time and location.

The contribution of this paper is twofold. First, we present a simple but general model for geographical load balancing and provide an offline formulation for solving the problem with deadline requirements. For each time slot, the formulation determines the assignment of workload to data centers and the migration of workload between data centers to adapt with the dynamic electricity price variation.

Second, we design an online algorithm for geographical load balancing considering migration and prediction error. The algorithm uses migration to improve the performance in case of prediction errors. We show that no online algorithm has constant competitive ratio with respect to the offline algorithm because of the uncertainty in electricity price variation. This allows us to compare our online algorithm with a simpler online algorithm without migration and prediction error and to determine a bound on the cost based on the prediction error. We then prove that an online algorithm with migration gives better cost savings than the online algorithm without migration, with future electricity price prediction. We validated our model by experiments using MapReduce traces as dynamic workload and found 20-30\% total cost savings.

The rest of the paper is organized as follows. Section II presents the model that we use to formulate the optimization and gives the offline formulation. In Section III, we present the online algorithm for determining workload assignment and migration dynamically for uniform and nonuniform deadline.
Section IV shows the experimental results. In Section V, we describe the state of the art research related to geographical load balancing and Section VI concludes the paper.

\section{Model Formulation}
In this section, we describe the model we use for geographical load balancing via dynamic deferral. The assumptions used in this model are minimal and this formulation captures many properties of current geographical load balancing and workload characteristics.

\subsection{Workload Model}
We consider a workload model where the total workload varies over time. The time interval we are interested in is $t\in \{0,1,\ldots,T\}$ where $T$ can be arbitrarily large. In practice, $T$ can be a year and the length of a time slot $\tau$ could be as small as milliseconds for service requests (e.g. HTTP) or as large as several minutes for batch-like jobs (e.g. MapReduce).  A basic assumption of our model is that energy (electricity) costs may vary in time, yet remain fixed within time slot length $\tau$. To facilitate the future price prediction, we denote the set of the time slots in a 24-Hour time frame by ${\cal K}\subset T$. In our model, the jobs have length less than $\tau$ and each job has deadline $D$ (in terms of number of slots) associated with it within which it needs to be executed where $D$ is a nonnegative integer. The value of $D$ can be zero for interactive jobs and large for batch-like jobs. If the length $\ell$ of a job is greater than $\tau$ then we can safely decompose it into small pieces ($\le \tau$) each of which is released after the execution of the preceding piece. If the job is preemptive then we assign deadline $\lfloor D/\ell\rfloor -1$ to each of the pieces, else for a non-preemptive job, we assign deadline of $D-\ell$ for the first piece and deadlines of zeros for the other pieces. Thus large jobs are decomposed into small jobs. Hence we do not distinguish each job, rather deal with the total amount of workload. First we consider the case of uniform deadlines, that is, deadline is uniform for all workloads, followed by non-uniform deadline case in Section~IIIE. Let $L_t$ be the amount of workload released at time slot $t$. This amount of work must be executed by the end of time slot $t+D$.

In our model, we consider a large computing facility (``cloud''), consisting of $n$ data centers. At each time $t$, the total workload $L_t$ arrive at a central dispatcher from which load balancing decisions are made. We assume that the arriving workload cannot be stored at the dispatcher, i.e., the workload arriving at the beginning of time $t$ needs to be dispatched to the data centers after the assignment of workload for each data center is determined. After the load balancing decisions are made at the dispatcher, the jobs can be stored at each data center to be executed at a suitable time before deadline. We are not concerned about the computation capability (homogeneous/heterogeneous) inside each data center rather we focus on load distribution considering data centers as computation units. The total computation capacity $M_i$ in data center $i$ is fixed and given for $1\le i\le n$. We normalize $L_t$ by the processing capability of the data centers i.e. $L_t$ denotes the computation units required to execute the workload at time $t$.

Let $x_{i,d,t}$ be the portion of the released workload $L_t$ that is assigned to be executed at data center $i$ at time slot $t+d$. Let $x_{i,t}$ be the total workload assigned to be executed at time $t$ to data center $i$ and $x_t$ be the total assignment at time $t$. Then $0\le x_{i,t}\le M_i$ and

$$\sum_{d=0}^{D} x_{i,d,t-d} = x_{i,t} \text{ and } \sum_{i=1}^{n} x_{i,t} = x_t$$

We assume that the energy prices vary unpredictably depending on time and location. The workload assigned to one data center can be migrated to other data centers in order to reduce the total energy consumption. Let $z_{i,j,d,t}$ be the amount of workload that is migrated at time $t$ from data center $i$ to be executed at data center $j$ at time $t+d$. Then $z_{i,j,t} = \sum_{d=0}^D z_{i,j,d,t}$, is the total amount of workload that is migrated form data center $i$ to $j$ at time $t$. We assume that there is a cost associated with each migration and all migrations are done as soon as the migration decisions are made. We also assume that migration time is negligible with regard to the time interval $\tau$ i.e. migration does not incur any delay. For service requests, $\tau$ is small as well as migration time is negligible and for batch-like jobs, $\tau$ is large in the range of minutes and migration time is in the range of seconds. Therefore with respect to $\tau$, migration time is small (negligible).

Since some portion of the assigned workload is migrated to other data centers, the workload that is executed at time $t$ at data center $i$ is the sum of the assigned workload and the net migrated-in workload as denoted by

\begin{equation}
\label{equn:execute}
y_{i,t} = x_{i,t} + \sum_{j=1}^n \sum_{d=0}^{D} z_{j,i,d,t-d} - \sum_{j=1}^n \sum_{d=0}^{D} z_{i,j,d,t-d}
\end{equation}

Since the released workload within $[1,T]$ needs to be finished within $T$ time slots, the total assignment and execution is equal to the total released workload over $T$ time slots as given by the following equation.

\begin{equation}
\label{equn:workloadconserv}
\sum_{t=1}^{T} \sum_{i=1}^{n}  y_{i,t} = \sum_{t=1}^{T} \sum_{i=1}^{n}  x_{i,t} = \sum_{t=1}^{T} L_{t}
\end{equation}

Thus there are two important decisions here: (i) determining $x_{i,d,t}$, assignment of workload to the data centers, and (ii) determining $z_{i,j,d,t}$, the amount of migrated workload during each time slot $t$.

\subsection{Cost Model}
The goal of this paper is to minimize the operation cost in the cloud which is the sum of the energy costs for executing workload at the data centers and the costs for migrating jobs between data centers.

\subsubsection*{Energy cost}
To capture the geographic diversity and variation of energy costs over time, we let $C_{i,t}(y_{i,t})$ denote the \emph{energy cost} for executing workload $y_{i,t}$ in data center $i$ at time slot $t$. We assume that $C_{i,t}(y_{i,t})$ is a nonnegative, (weakly) convex increasing function as used in \cite{Buchbinder}. Note that the function itself can change with time, which allows for time variation in energy prices.
The simplest example for the cost function for a time slot is an affine function which is the common model for the energy cost for typical data centers:
$$C_{i,t}(y_{i,t})=\alpha_i + \beta_{i,t} y_{i,t}$$

where $\alpha_i$ and $\beta_{i,t}$ are constants for data center $i$ and time slot $t$ (e.g. see \cite{GLB_Wierman}) and $y_{i,t}$ is the executed workload to data center $i$ at time $t$. Note that in this model, the load dependent component of the function only depends on time, because the real time electricity price varies with real time demand.

\subsubsection*{Migration cost}
The \emph{migration cost} is the cost for migrating workload from one data center to another which accounts for bandwidth cost and energy consumed by the intermediate devices. The migration cost is proportional to the amount of migrated workload which is represented by,
$$B_{i,j}(z_{i,j,t}) = b_{i,j}z_{i,j,t}$$

where $b_{i,j}$ is constant for migration from data center $i$ to $j$ regardless of the migration time \cite{Buchbinder}. Note that there are also bandwidth costs associated with the arrival of jobs into the cloud and leaving the cloud. But these costs can be easily incorporated without changing the problem formulation, as they are constant and do not depend on the migration control.

\subsection{Optimization Problem}
Given the models above, the goal of geographical load balancing is to choose the migrating jobs $z_{i,j,d,t}$ and the dispatching rule $x_{i,d,t}$ to minimize the total cost during $[1,T]$, which is captured by optimization~(\ref{equn:opt1}). In this formulation, constraint (\ref{equn:opt1}b) represents that the total assignment should be equal to the total released workload and constraint (\ref{equn:opt1}c) represents that the total workload that is migrated from data center $i$ to be executed at time $t$ cannot exceed the assigned workload to be executed at time $t$ at data center $i$. We now prove that there is an optimal solution of optimization (\ref{equn:opt1}) where there is no migration. We have the following lemma.

\begin{figure*}
\begin{IEEEeqnarray}{lll}
\label{equn:opt1}
 \min_{x_t,z_t}& \sum_{t=1}^T \sum_{i=1}^{n}  C_{i,t}(y_{i,t}) +  \sum_{t=1}^T \sum_{i=1}^{n} \sum_{j=1}^{n} b_{i,j} z_{i,j,t} \quad & \IEEEyessubnumber\\
 \text{subject to} \quad &  \sum_{i=1}^{n} \sum_{d=0}^{D} x_{i,d,t} = L_t  & \forall t   \IEEEyessubnumber\\
 &  \sum_{k=0}^{D-d} \sum_{j=1}^{n} z_{i,j,d+k,t-k} \le \sum_{k=0}^{D-d} x_{i,d+k,t-k}& \forall i, \forall d, \forall t  \IEEEyessubnumber\\
 &  y_{i,t} \le M_i & \forall i, \forall t  \IEEEyessubnumber\\
 &  x_{i,d,t}\ge 0, z_{i,j,d,t}\ge 0 & \forall i, \forall j, \forall d, \forall t.  \IEEEyessubnumber
\end{IEEEeqnarray}
\end{figure*}

\begin{lemma}
\label{lemma:nomigration}
In every optimal solution of optimization (\ref{equn:opt1}), either $z_{i,j,d,t} = 0$,  for all time slots $t$, and deferral $d$ or $b_{i,j}=0$, for all $(i, j)$.
\end{lemma}

\begin{IEEEproof}
Suppose for a contradiction that the optimal solution $O$ contains $z_{i,j,d,t}>0$, and $b_{i,j}>0$, for some $(i,j)$. Then we can construct another optimal solution $O'$ where $z'_{i,j,d,t}=0$, $\forall i,j,d,t$ by making $x'_{i,t}=y_{i,t}$. Then $y'_{i,t}=y_{i,t}$, $\forall i,t$ and $\sum_{t=1}^T \sum_{i=1}^{n}  C_{i,t}(y_{i,t}) = \sum_{t=1}^T \sum_{i=1}^{n}  C_{i,t}(y'_{i,t})$. The objective value of the solutions $obj(O')<obj(O)$ because $\sum_{t=1}^T \sum_{i=1}^{n} \sum_{j=1}^{n} b_{i,j} z'_{i,j,t} = 0 < \sum_{t=1}^T \sum_{i=1}^{n} \sum_{j=1}^{n} b_{i,j} z_{i,j,t}$ since $b_{i,j} > 0$. This contradicts the assumption that $O$ is an optimal solution.
\end{IEEEproof}

\begin{corollary}
\label{cor:nomigration}
There exists an optimal solution of optimization (\ref{equn:opt1}) where $z_{i,j,d,t} = 0$,  $\forall i, j, d, t$.
\end{corollary}

By Corollary~\ref{cor:nomigration}, migration is unnecessary when all the information about workload and energy price are known in advance. However, when all the information are not available then migration becomes important as investigated in the next section.

Since the operating cost $C_{i,t}(\cdot)$ is an affine function, the objective function is linear as well as the constraints. Hence it is clear that the optimization~(\ref{equn:opt1}) is a linear program. Note that the workload $x_{i,t}$ in the formulation is not considered to be integer. This is acceptable because the number of requests in data centers at each time slot is in the range of thousands and we can round the resulting assignment with minimal increase in cost. If all the future costs and workload were known in advance, then the problem could be optimally solved as a linear program. However our basic assumption is that electricity prices change in an unpredicted manner depending on time and location. Therefore, we tackle the optimization problem as an online optimization problem.

\section{Online Algorithm}
In this section we consider the online case, where at any time $t$, we neither have information about the future workload $L_{t'}$ for $t'>t$, nor have knowledge about future electricity prices. The workload released at time $t$ can be delayed to be executed in future time slots if the cost for execution at future time slots is less than the current cost. We apply optimization on the current and delayed workload and distribute them in future time slots so that the total cost for execution and migration is minimized subject to the future predicted prices. In the online algorithm, we decouple the migration decision from the assignment decision and apply optimization in two levels: (i) dispatcher level and (ii) data center level. The dispatcher makes decision about the assignment of the incoming workload to data centers based on predicted future electricity prices. Then the data centers make decision on adjusting the execution of the workload in current and future time slots and the migration of workload between data centers in case of prediction errors.

\subsection{Electricity Price Prediction Model}
In this section we illustrate our model for predicting future price of electricity. Since only the load proportional component of energy cost depends on time, we need to predict $\tilde{\beta}_{i,t}$. Then the predicted energy cost function will be $$\tilde{C}_{i,t}(y_{i,t})=\alpha_i + \tilde{\beta}_{i,t} y_{i,t}$$

The load dependent electricity price $\beta_{i,t}$ is announced by the utility at location $i$ at the beginning of each time slot $t$ and is kept constant during the duration of that time slot. However future prices will change independently of past prices according to some known probability density function. Predicting electricity prices is difficult because price series present such characteristics as nonconstant mean and variance and significant outliers. We model the prediction noise by a Gaussian random variable with zero mean and variance to be estimated. In other words, we model future prices within a 24-hour time-frame by Gaussian random variables with known means, which are the predicted prices, and some estimated variance. The mean for the Gaussian distribution is predicted by the widely used moving average method for time series. The  variance for the Gaussian distribution is estimated from the history by the weighted average price prediction filter proposed in \cite{mohsenian_residential}.  In this model, variances are predicted by linear regression from the previous prices from yesterday, the day before yesterday and the same day last week. By using two different methods for mean and variance, we exploit both the temporal and historical correlation of electricity prices. Let $\tilde{\mu}^\kappa_i[\chi]$ and $\tilde{\sigma}_i^\kappa[\chi]$ be the predicted means and standard deviations for each time slot $\kappa$ on day $\chi$ for geographical location $i$. Then the mean of the prediction model for Gaussian distribution is obtained as follows:

$$\tilde{\mu}_i^\kappa = \varepsilon_0 + \sum_{j=0}^D  \varepsilon_{\kappa-j} \beta_{i,\kappa-j}, \quad \forall i\in n, \forall \kappa \in \cal{K}$$

Here, $\varepsilon_j$ are the coefficients for the moving average method which can be estimated by training the model over the previous day prices. The variance parameter  $\tilde{\sigma}_i^\kappa[\chi]$ is estimated from the history using the following equation:

\begin{eqnarray*}
\tilde{\sigma}_i^\kappa[\chi] = k_1 {\sigma}_i^\kappa[\chi-1]  + k_2 \sigma_i^\kappa[\chi-2]  + k_7 {\sigma}_i^\kappa[\chi-7], \\ \quad \forall i\in n, \forall \kappa \in \cal{K}
\end{eqnarray*}

Here, ${\sigma}^\kappa_i[\chi-1]$, ${\sigma}_i^\kappa[\chi-2]$ and ${\sigma}_i^\kappa[\chi-7]$ denote the previous standard deviation values ${\sigma}^\kappa_i$ on yesterday, the day before yesterday and the same day last week, respectively. The coefficients for the weighted average price prediction filter $k_1$, $k_2$ and $k_7$ are selected from \cite{mohsenian_residential}.

\subsection{Optimization for Dispatcher}
The dispatcher makes decision on the assignment of the workload to the data centers based on the current electricity prices and future price predictions. The following optimization applied at the dispatcher determines the assignment of workload $x_{i,d,t}$ to data centers for $0\le d \le D$ and $1\le i\le n$.

\begin{IEEEeqnarray}{lll}
\label{equn:opt3}
 \min_{x_{i,d,t}} \quad &  \sum_{i=1}^{n}\sum_{d=0}^{D} C_{i,t}(x_{i,d,t-d}) \nonumber\\ & + \sum_{i=1}^{n} \sum_{k=1}^{D} \sum_{d=k}^D \tilde{C}_{i,t+k}(x_{i,d,t+k-d})& \quad \IEEEyessubnumber \\
 \text{subj. to} \quad & \sum_{i=1}^{n} \sum_{d=0}^{D} x_{i,d,t} = L_t &  \IEEEyessubnumber\\
 &  0\le \sum_{d=s-t}^{D} x_{i,d,s-d} \le M_i \quad   \forall i, t\le s\le t+D.&  \quad  \IEEEyessubnumber
\end{IEEEeqnarray}

where $\tilde{C}_{i,t'}()$ is the predicted cost function at time ${t'>t}$ for data center $i$ and $x_{i,d,t''}$ is the unexecuted workload at data center $i$ that was assigned at time $t''<t$ to be executed at time $t''+d$ where $ t-t''\le d \le D$. Note that greedy method can also be applied to compute the optimum assignment for the dispatcher.

\subsection{Optimization for Data Centers}
The predicted electricity prices at time $t$ may contain prediction errors which may lead to some badness in the assignment. Data centers can migrate workload between each other to adjust the assignment to overcome prediction errors for minimizing the total cost in the later time slots. The adjustment is made by applying an optimization on the schedule for the unexecuted workload for the current and future time slots. For each data center $i$, this optimization makes decision on how much workload to execute at time $t$, how much to defer to execute later and how much to migrate to other data centers. Note that the workload released at or before $t$, cannot be delayed to be assigned after time slot $t+D$. Hence we minimize the total cost by applying optimization on the already released but unexecuted (delayed) workload over the interval $[t,t+D]$. We have two versions of the online optimization at data centers. First we formulate the optimization without considering migration. Then the more general case with migration is considered.

\subsubsection{Formulation without Migration}
\begin{figure}[!t]
\begin{center}
\includegraphics[width=2.8in]{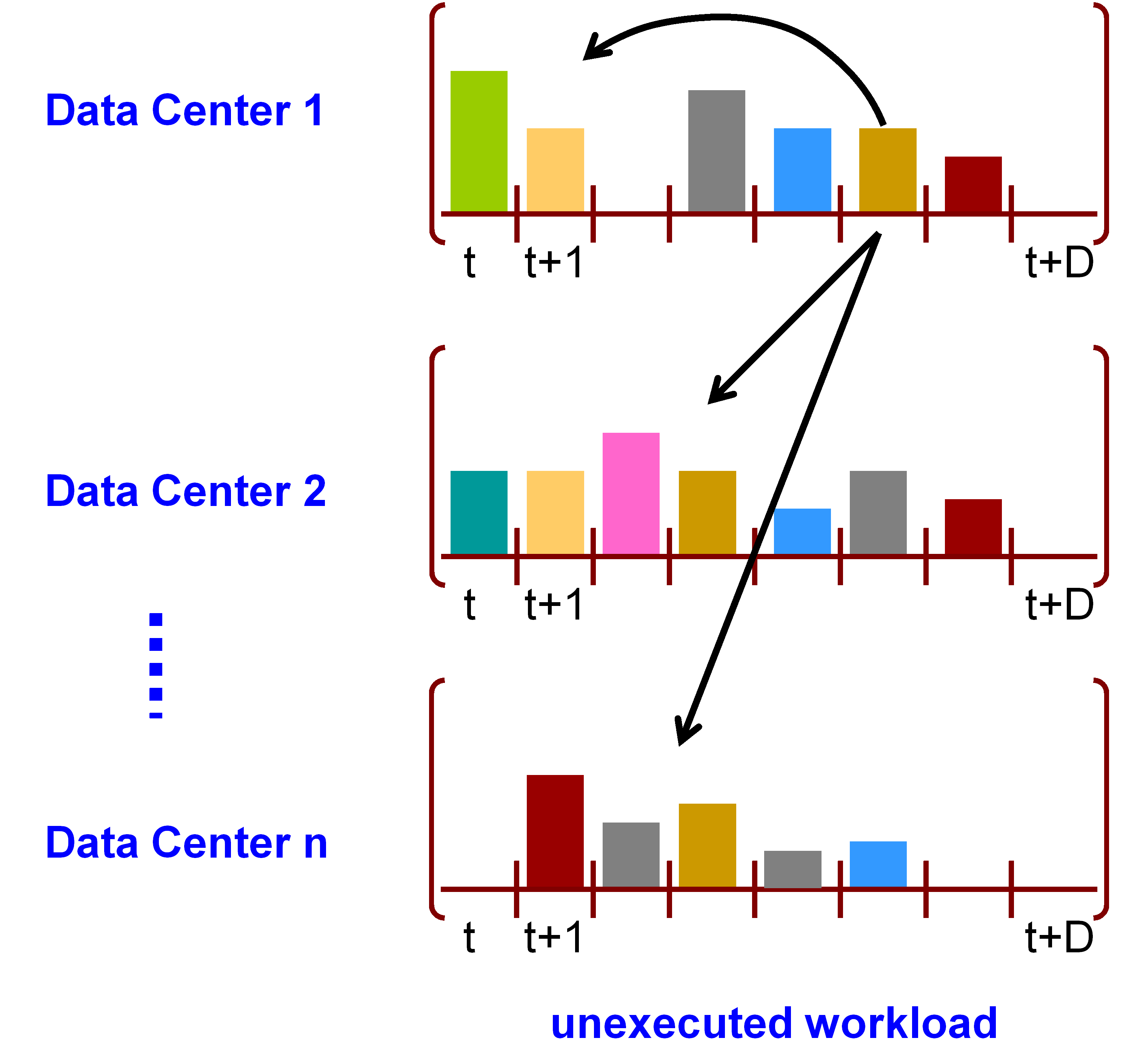}
\caption{Optimization on unexecuted (assigned) workload at time $t$ by transferring workload to previous time slots (curved arrow) and by migration to other data centers (straight arrow).}
\label{fig:unexecuted}
\end{center}
\end{figure}

We start with a formulation for the online case by considering load balancing without migration. Although there is no migration, still the data centers can improve the assignment by executing the delayed workload early in previous time slots without violating deadline as shown by the curved arrow in Figure~\ref{fig:unexecuted}. Let $u_{i,t}$ and $w_{i,t}$ denote the assigned (delayed) and executed workload at time $t$ at data center $i$, respectively. Initially $u_{i,t} = w_{i,t} = 0$, $\forall t$. Then the values of $w_{i,s}$ for $t\le s \le t+D$ are obtained at each time $t$ by applying the optimization~(\ref{equn:opt4u}). And the values of $u_{i,s}$ for $t\le s \le t+D$ are updated each time $t$ from the computed $w_{i,s}$. The following optimization determines the current and future execution variables $w_{i,s}$ for $t\le s \le t+D$.

\begin{IEEEeqnarray}{lll}
\label{equn:opt4u}
 \min_{w_{i,t}}\quad & \sum_{i=1}^{n}  C_{i,t}(w_{i,t}) + \sum_{i=1}^{n} \sum_{s=t+1}^{t+D} {\tilde{C}}_{i,s}(w_{i,s}) & \IEEEyessubnumber\\
 \text{subj. to}\quad & \sum_{s=t}^{t+D} w_{i,s} = \sum_{s=t}^{t+D} u_{i,s} \quad \quad \forall i &  \IEEEyessubnumber\\
 & \sum_{r=t}^{s} w_{i,r} \ge \sum_{r=t}^{s} u_{i,r} \quad \quad \forall i, t\le s< t+D & \IEEEyessubnumber\\
 &  0 \le w_{i,s} \le M_i \quad \quad \quad \quad \forall i, t\le s\le t+D. & \IEEEyessubnumber
\end{IEEEeqnarray}

Here the constraints (\ref{equn:opt4u}b) and (\ref{equn:opt4u}c) ensure that the assigned workload $u_{i,s}$ can only be moved to an earlier time slot $t\le t'\le s$ and thus does not violate deadline.

\subsubsection{Formulation with Migration}
We can utilize the migration of workload between data centers to correct the prediction errors (to some extent) made by the dispatcher at an early slot during dispatching. Let $z_{i,j,d,t}$ denote the migration of workload from data center $i$ at time $t$ which will be executed at data center $j$ at time $t+d$. Then the values of $u_{i,s}$ for $t\le s \le t+D$ are updated each time $t$ from $w_{i,t}$, $z_{i,j,d,t}$ and $z_{j,i,d,t}$'s with the rule in equation (\ref{equn:u_migration}).

\begin{figure*}[!htbp]
\begin{equation}
\label{equn:u_migration}
u_{i,s} =
\begin{cases}
x_{i,D,t} & \text{if }s=t+D,\\
w_{i,s} + x_{i,s-t,t} + \sum_{j=1}^n z_{j,i,s-t+1,t-1} - \sum_{j=1}^n z_{i,j,s-t+1,t-1}  \qquad &   \text{if } t\le s < t+D.
\end{cases}
\end{equation}
\end{figure*}

\begin{figure*}[!htbp]
\begin{IEEEeqnarray}{llll}
\label{equn:opt4}
 \min_{w_{i,t},z_{i,j,d,t}} \quad & \sum_{i=1}^{n}  C_{i,t}(y_{i,t}) + \sum_{i=1}^{n} \sum_{s=t+1}^{t+D}   {\tilde{C}}_{i,s}(y_{i,s}) +  \sum_{i=1}^{n} \sum_{j=1}^{n} \sum_{d=0}^{D}  b_{i,j} z_{i,j,d,t} & \IEEEyessubnumber\\
 \text{subject to} \quad & \sum_{i=1}^n \sum_{s=t}^{t+D} y_{i,s} = \sum_{i=1}^n \sum_{s=t}^{t+D} u_{i,s} &   \IEEEyessubnumber\\
 & \sum_{i=1}^n \sum_{r=t}^{s} y_{i,r} \ge \sum_{i=1}^n \sum_{r=t}^{s} u_{i,r} & t\le s< t+D  \IEEEyessubnumber\\
 &  \sum_{j=1}^{n} z_{i,j,s-t,t} \le  w_{i,s}  & \forall i, t\le s\le t+D  \IEEEyessubnumber\\
 &  y_{i,s} \le M_i, w_{i,s}\ge 0  & \forall i, t\le s\le t+D  \IEEEyessubnumber\\
 &  z_{i,j,d,t}\ge 0  & \forall i, \forall j, \forall d.  \IEEEyessubnumber
\end{IEEEeqnarray}
\end{figure*}

Then applying the optimization~(\ref{equn:opt4}), the values for $w_{i,s}$ and $z_{i,j,d,s-d}$ are determined for $t\le s \le t+D$. Then the workload that is executed including the migration at data center $i$ at time $t$ is $y_{i,s}$, for $t\le s\le t+D$, which is determined by the following equation

\begin{equation*}
\label{equn:execute_datacenter}
y_{i,s} = w_{i,s} + \sum_{j=1}^n \sum_{d=0}^{D} z_{j,i,d,s-d} - \sum_{j=1}^n \sum_{d=0}^{D} z_{i,j,d,s-d}
\end{equation*}

The optimization~(\ref{equn:opt4}) applied at time $t$, determines the current and future execution variable $w_{i,s}$ for $t\le s \le t+D$ and the current migration $z_{i,j,d,t}$ for $0\le d\le D$.

Here the constraints (\ref{equn:opt4}b) and (\ref{equn:opt4}c) ensure that the assigned workload $u_{i,s}$ can be migrated to other data centers and can only move to an earlier time slot $t\le t'\le s$ and thus does not violate deadline as shown by arrows in Figure~\ref{fig:unexecuted}. Constraint (\ref{equn:opt4}d) ensures that the amount of migration does not exceed the unexecuted workload. Then the actual workload that is executed at time $t$ at data center $i$ is,

\begin{equation}
\label{equn:execworkload}
y_{i,t} = w_{i,t} + \sum_{j=1}^{n}  z_{j,i,0,t} - \sum_{j=1}^{n} z_{i,j,0,t}
\end{equation}

In summary, at the beginning of each time slot, we apply the optimization (\ref{equn:opt3}) at the dispatcher and then the tasks are assigned to the data centers. Then the optimization (\ref{equn:opt4}) for the data centers, is applied globally to determine the assignment and migration of previously released unexecuted workload. Then the migration takes place and the amount of execution for each data center is determined by equation (\ref{equn:execworkload}). After that each of the data centers execute that amount of workload.

\subsection{Analysis of the Algorithm}
We now analyze the performance of the online algorithm. We first prove that there does not exist any online algorithm with constant competitive ratio with respect to the offline formulation~(\ref{equn:opt1}).


\begin{lemma}
\label{lemma:online2}
No online algorithm has constant competitive ratio with respect to the offline formulation~(\ref{equn:opt1}).
\end{lemma}

\begin{IEEEproof}
Prediction error degrades the performance of the online algorithm, hence w.l.o.g. we assume that the online algorithm does not have any prediction error i.e. $\epsilon = 0$. We prove the claim by adversary method i.e. we  consider  an  adversary  who presents the online  algorithm  with several  different  instances.  Suppose we have only one data center $n=1$ with capacity $M$. The time slots are $\{0,1,\ldots, T\}$.  The uniform deadline is $D<T$. And the cost function parameters $\beta_t$ are $\beta_0 = K\cdot \beta_D$ and $\beta_t \gg K\cdot \beta_D$ for $t\in\{0,1,\ldots, T\} - \{0,D\}$. Now for determining the assignments $x_{d,t}$, we consider two cases:

\Case 1 \{$x_{D,0}\neq 0$\}\\
In this case, suppose the workload released at time $t=0$ and $t=1$ are $L_0$ and $L_1$ respectively and $L_0=L_1=M$. Then the online assignment vector has $x_{d,1}> 0$ for some $0\le d\neq D-1$ where $\beta_{d+1}\gg K\cdot \beta_D$ which can be arbitrarily large. Hence the competitive ratio becomes unbounded.

\Case 2 \{$x_{D,0}=0$\}\\
In this case, we construct an adversary input by making $L_0=M$ and $L_1 = 0$. The offline algorithm chooses $x^*_{D,0} = L_0$ but the online algorithm chooses $x_{0,0}=L_0$. Since $\beta_0 = K.\beta_D$, the competitive ratio for this case is $\sim K$. Since $K$ is arbitrary, the competitive ratio is not bounded by a constant.
\end{IEEEproof}

Since the performance of any online algorithm cannot be bounded with respect to offline algorithm, we compare the online algorithm with a simple online algorithm without migration and without any prediction error. We call such an online algorithm as $A$ and the online algorithm with migration (described in Section IIIC2) as $A^m_\epsilon$. We denote the online algorithm with prediction error but without migration (described in Section IIIC1) as $A_\epsilon$. Basically we are going to compare the performance of $A$ and $A^m_\epsilon$. We denote the total cost from an algorithm $A$ as $cost(A)$. We have the following lemma.

\begin{lemma}
\label{lemma:online22}
$cost(A^m_\epsilon) \le cost(A_\epsilon)$.
\end{lemma}

\begin{IEEEproof}
Let $y^m_{i,t}$ and $y_{i,t}$ be the workload executed at time $t$ by algorithms $A^m_\epsilon$ and $A_\epsilon$ respectively. In the algorithm $A^m_\epsilon$, the workload assigned to a time slot $t$ can only move to earlier time slot $t'\le t$ as illustrated by constraints (\ref{equn:opt4}b) and (\ref{equn:opt4}c). Hence $\sum_{i=1}^n y_{i,t} \le \sum_{i=1}^n y^m_{i,t}$. Therefore $\Delta y = \sum_{s=t+1}^{t+D} \sum_{i=1}^n y^m_{i,s} - \sum_{s=t+1}^{t+D} \sum_{i=1}^n y_{i,s}\ge 0$. That means we have $\Delta y$ more workload to execute in later slots $s>t$ for $A_\epsilon$ than $A^m_\epsilon$ and due to optimization at $A^m_\epsilon$, $\sum_{i=1}^n C_{i,t}(\Delta y) + \sum_{i=1}^n \sum_{j=1}^n b_{i,j} (\Delta y) \le \sum_{i=1}^n \tilde{C}_{i,s}(\Delta y)$ for any $t+1\le s\le t+D$. Since both the algorithms use the same energy cost functions, we have $cost(A^m_\epsilon) \le cost(A_\epsilon)$.
\end{IEEEproof}

According to Lemma~\ref{lemma:online22}, incorporating migration into the online algorithm reduces the total cost of execution than $A_\epsilon$. Using this lemma, we now bound the cost for algorithm $A_\epsilon^m$ with respect to $A$ by the prediction error $\epsilon$ as stated in the following theorem.

\begin{theorem}
\label{theorem:online3}
$cost(A^m_\epsilon)\le (1+\epsilon)\cdot cost(A)$.
\end{theorem}

\begin{IEEEproof}
We first show that $cost(A_\epsilon)\le (1+\epsilon)\cdot cost(A)$. Then by lemma~\ref{lemma:online22}, the theorem holds. If there were no prediction error i.e. $\epsilon=0$, $cost(A_\epsilon) = cost(A)$. If there is a prediction error $\epsilon>0$ then suppose the electricity price predicted for time $t$ by $A_\epsilon$ is $\tilde{\beta}$, whereas the actual price used in $A$ is $\beta$. Then $\tilde{\beta}-\epsilon \le \beta \le \tilde{\beta}+\epsilon$. Then $\frac{cost(A_\epsilon)}{cost(A)}= \frac{\alpha+\tilde{\beta}y}{\alpha+{\beta}y} \le 1+\frac{\epsilon}{\beta}\le (1+\epsilon)$, where $y$ is the executed workload.
\end{IEEEproof}

Suppose the prediction error follows the Gaussian distribution with standard deviation $\sigma$. Then the probability that the prediction error is bounded by $\epsilon$ is given by the Chebyshev's inequality
$$Pr(|C-\tilde{C}|\ge \epsilon)\le \frac{\sigma^2}{\epsilon^2}$$

By Theorem~\ref{theorem:online3}, the cost savings from the online algorithm depends on the price variation and the quality of prediction.

\subsection{Nonuniform Deadline}
The algorithm described above can be easily extended for nonuniform deadline where the deadline requirement is not same for all the workload. In this case the workload can be decomposed according to their associated deadline. Suppose $L_{d,t}\ge 0$ be the portion of the workload released at time $t$ and has deadline $d$, for $0\le d\le D$, where $D$ is the maximum deadline. Then we have $$\sum_{d=0}^{D} L_{d,t} = L_t.$$ Then the constraints for $L_t$ in the offline formulation~(\ref{equn:opt1}b) and the online formulation~(\ref{equn:opt3}b) can be replaced by the following constraint:
$$\sum_{i=1}^n \sum_{k=0}^d x_{i,k,t} = \sum_{k=0}^d L_{k,t}, \qquad 0\le d\le D$$
Then the same algorithm can be applied to get solutions for nonuniform deadline.

\section{Experimental Results}
In this section, we seek to evaluate the cost incurred by the algorithms $A$, $A_\epsilon$ and $A_\epsilon^m$ relative to the optimal solution in the context of workload generated from realistic data.


\subsection{Experimental Setup}
We aim to use realistic parameters in the experimental setup and provide conservative estimates of the cost savings resulting from optimal geographical load balancing.

\subsubsection*{Electricity Price}
There are two types of electricity markets: Wholesale Market and Retail Market. Due to the high consumption of electricity in data centers, they usually purchase electricity from the wholesale markets \cite{GLB_Wierman}. Electricity price varies on a 5 minute or 15 minute basis in real time wholesale electricity market. Electricity price in this market exhibit significant volatility with high frequency variation \cite{Qureshi}.

\begin{figure}[!t]
\begin{center}
\includegraphics[width=3.5in]{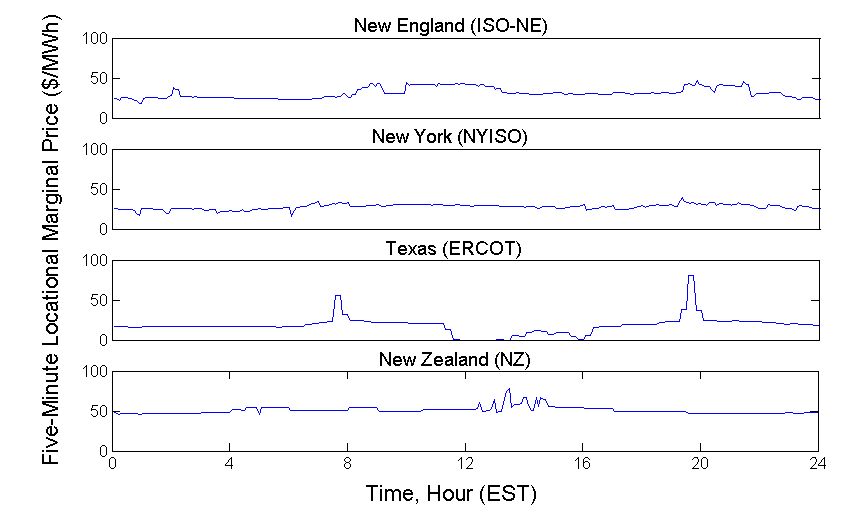}
\caption{Illustration of five minute locational marginal electricity prices in real time market on 15th February, 2012 for four different regions (a) New England (ISO-NE), (b) New York (NYISO), (c) Texas (ERCOT), (d) New Zealand (NZ).}
\label{fig:price_day}
\end{center}
\end{figure}

We run our simulations for four data centers geographically located in four different locations. We choose distant locations for our experiments. We choose the locations near those power grids whose real time electricity prices are publicly available. We used the publicly available data from electricity markets from Independent System Operator New England (ISO-NE) \cite{ne_iso}, New York Independent System Operator (NYISO) \cite{nyiso}, Electric Reliability Council of Texas (ERCOT) \cite{ercot} and Electricity Market of New Zealand (NZ) \cite{nz}. We took the locational based marginal prices (LBMP) from the 5 minute spot markets for three days (15th, 14th and 8th February, 2012) and ran our experiments on the prices of 15th February using the prices for 14th and 8th for prediction of future prices. We use the four locations to have both temporal and geographical variation of electricity prices e.g. the time zones of New York, New England, Texas and New Zealand are GMT-5, GMT-7, GMT-6 and GMT+13 respectively. The variation of electricity prices for different locations are plotted in Figure~\ref{fig:price_day} with Eastern Standard Time (EST). These graphs indicate significant spatio-temporal variation in electricity prices.

\subsubsection*{Workload Description}
\begin{figure}[!t]
\centerline{\subfigure[Workload A]{\includegraphics[width
=1.7in]{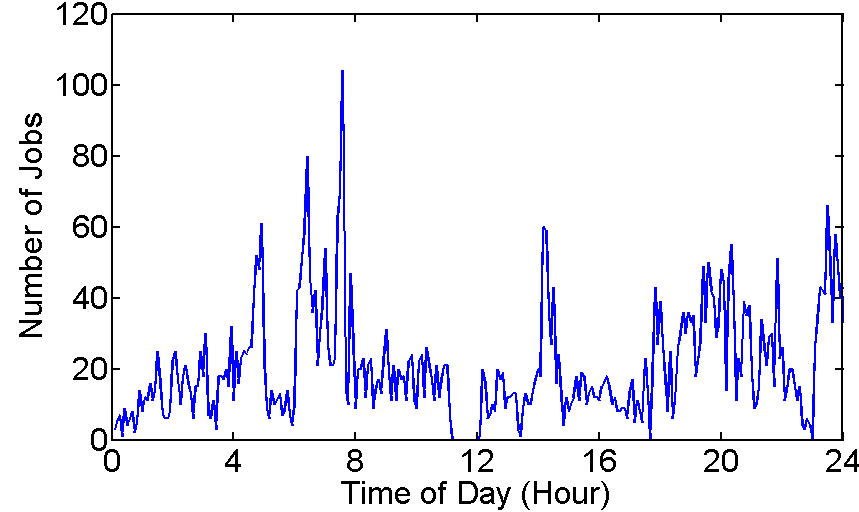}
}\hfil
\subfigure[Workload B]{\includegraphics[width=1.7in]{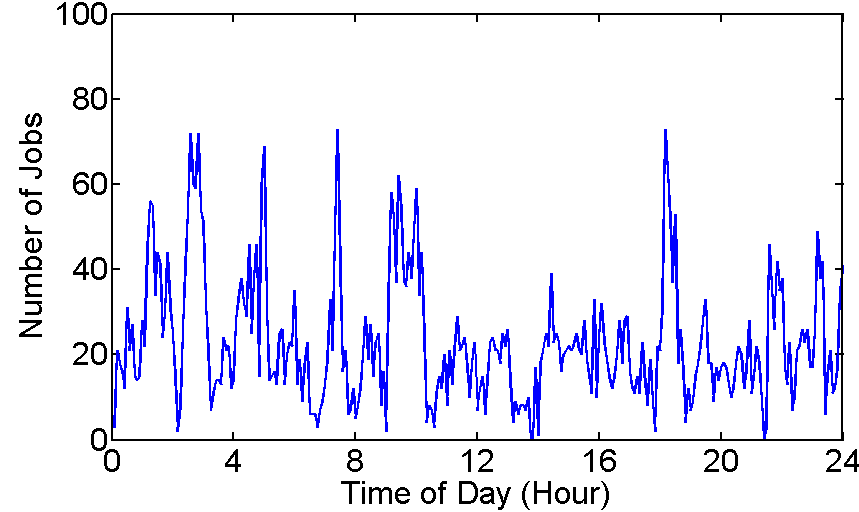}
}}
\caption{Illustration of the traces for dynamic workload used in the experiments.}
\label{fig:workload}
\end{figure}

We use two publicly available MapReduce traces as examples of dynamic workload. The MapReduce traces were released by Chen et al. \cite{n4} which are produced from real Facebook traces for one day (24 hours) from a cluster of 600 machines. We count the number of different types of job submissions over a time slot length of 5 minutes and use that as a dynamic workload (Figure~\ref{fig:workload}) for simulation. The two samples we use, represent strong diurnal properties and have variation from typical workload (Workload A) to bursty workload (Workload B).

We use time slot length of 5 minutes because the electricity prices vary with an interval of 5 minutes. In practice, load balancing decisions can be made more frequently with slot length size in the range of seconds. We then assign deadline for each job in terms of the number of slots the job can be delayed. For the case of uniform deadline, we vary deadline $D$ from $1-12$ for the simulation. This is realistic because MapReduce workloads have deadlines in the range of minutes as deadlines from 8-30 minutes for these workloads have been used in the literature \cite{n9,n10,n11}. For the non-uniform case, we use k-means clustering to classify the MapReduce workload into 10 groups based on the total sizes of map, shuffle and reduce bytes. The characteristics of each group are depicted in Table~\ref{table:nonuniform} where smaller jobs dominate the workload mix, as smaller jobs form larger classes and larger jobs form smaller classes. This kind of clustering has been used by Chen et al. for classifying the workload. For each class of jobs, we assign a deadline from $1-10$ slots such that smaller class (batch jobs) has larger deadline and larger class (interactive jobs) has smaller deadline.

\begin{table}[!ht]
\caption{Cluster Sizes and Deadlines for Workload Classification by k-means clustering for Nonuniform Deadline} 
\centering  
\begin{tabular}{c |c| r| c| r| c} 
\hline\hline                        
Cluster & \multicolumn{2}{|c|}{Workload A}  & \multicolumn{2}{|c|}{Workload B}  & Deadline  \\ 
\cline{2-3}  \cline{3-5}
 & \#Jobs & \multicolumn{1}{|c|}{GB} & \#Jobs & \multicolumn{1}{|c|}{GB}& \#slots \\
\hline                  
1 & 4878 & 0.22 & 5632 & 0.32 & 1\\ 
2 & 496 & 3.13 & 513 & 5.85 & 2\\
3 & 196 & 9.90 & 170 & 18.62 & 3\\
4 & 113 & 23.49 & 100 & 39.09 & 4\\
5 & 80 & 49.59 & 106 & 56.52 & 5\\ 
6 & 49 & 85.07 & 44 & 99.23 & 6\\
7 & 48 & 146.67 & 26 & 160.90 & 7\\
8 & 19 & 286.36 & 29 & 350.62 & 8\\
9 & 13 & 620.01 & 11 & 659.30 & 9\\
10 & 2 & 8104.52 & 7 & 1294.19 & 10\\
\hline 
\end{tabular}
\label{table:nonuniform} 
\end{table}

\subsubsection*{Cost benchmark}
Currently geographical load balancing for data centers typically does not use deferral of workload for load balancing \cite{GLB_Wierman, Buchbinder}. Often the load balancing decisions are made dynamically using greedy method based on current electricity prices without dynamic deferral. Clearly we could be energy efficient if we consider deferral of some of the tasks and use migration to adapt with the variation of electricity prices. We compare the total cost from the offline and online ($A_\epsilon^m$) algorithms with the greedy strategy as proposed by Qureshi et al. \cite{Qureshi} and evaluate the cost reduction. We also compare the total cost for the online algorithms $A_\epsilon$ and $A_\epsilon^m$.

\subsubsection*{Cost function parameters}
The cost function parameter $\beta_{i,t}$ is determined using current electricity price and $\tilde{\beta}_{i,t'}$, for $t'>t$, are determined using the electricity price prediction models described in Section IIIA. For our simulations, we use load independent parameter $\alpha_{i}=0$, for all $i$. The values for $b_{i,j}$ are determined proportional to the geographic distance between data centers $i$ and $j$. Since the workload cannot be migrated from source to source, we use $b_{i,i}$ to be a large number. Depending on the nature of the workload we varied the total capacity of the data centers because the algorithms keep on assigning the workload to the data center with the lowest cost until the data center is overloaded. Choosing a maximum capacity value to be less than the peak value allows us to visualize the cut off for the assignment. For both the workload, we use capacity $M_i = 50$ for all $i$. 

The future electricity prices $\tilde{\beta}_{i,t}$  for the next $D$ time slots are randomly generated from Gaussian Distributions because of their high unpredictability and the volume ($D$) of generation as described in Section IIIA. We use the same mean but different variances for the generation in each time slot. We use the optimal daily coefficients for the price prediction filter from \cite{mohsenian_residential} for estimating $\tilde{\sigma}_i^\kappa[\chi]$. Since we use the electricity prices for Wednesday (15th February, 2012), we choose $k_1 = 0.837$, $k_2 = 0$ and $k_7 = 0.142$. For the previous standard deviation values ($\sigma_i^\kappa[\chi-1]$, $\sigma_i^\kappa[\chi-7]$), we use the past standard deviation of electricity prices for $D$ slots on those days such that $\sigma_i^\kappa[\chi-1] := std(\beta_{i,\kappa}[\chi-1], \beta_{i,\kappa-1}[\chi-1], \ldots, \beta_{i,\kappa-D}[\chi-1])$ and $\sigma_i^\kappa[\chi-7] := std(\beta_{i,\kappa}[\chi-7], \beta_{i,\kappa-1}[\chi-7], \ldots, \beta_{i,\kappa-D}[\chi-7])$; where $std(\cdot)$ denotes the standard deviation. The mean $\tilde{\mu}_i^\kappa[\chi]$ is computed from the moving average of the prices for $D$ previous slots on the current day $\chi$.

\subsection{Experimental Analysis}
We now evaluate and analyze the cost savings provided by the offline and online algorithms.

\subsubsection*{Uniform Deadline}
\begin{figure}[!t]
\centerline{\subfigure[Workload A]{\includegraphics[width
=1.7in]{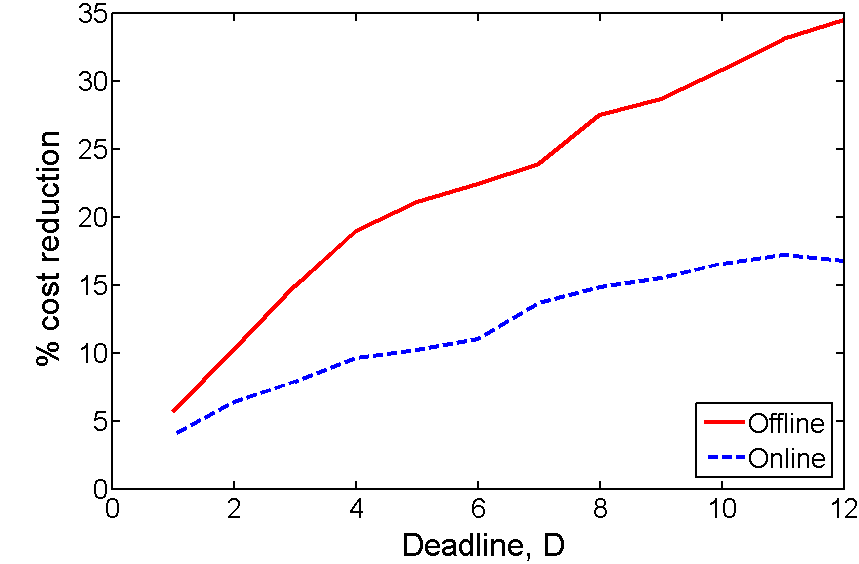}
}
\hfil
\subfigure[Workload B]{\includegraphics[width=1.7in]{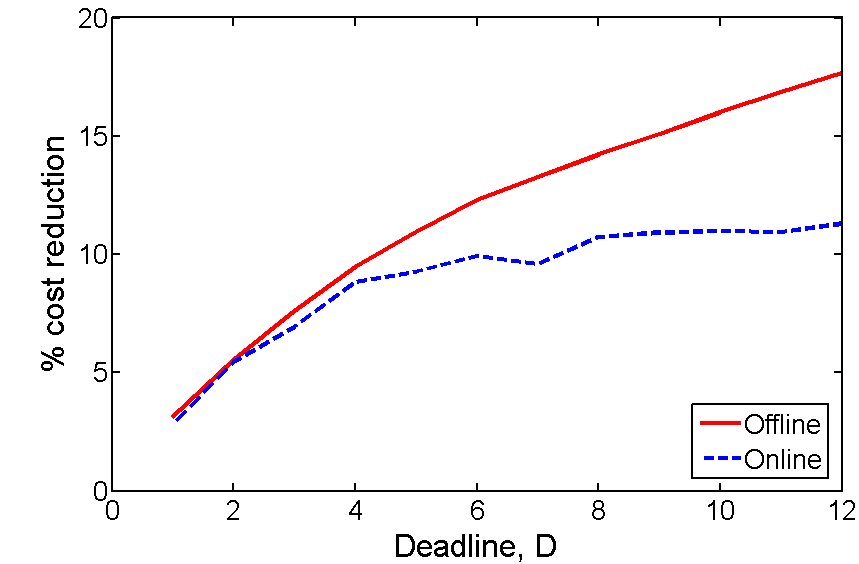}
}}
\caption{Impact of deadline on cost reduction by the offline and the online algorithm $A_\epsilon^m$ in comparison to greedy algorithm.}
\label{fig:cost_reduction}
\end{figure}

\begin{figure}[!t]
\centerline{\subfigure[Workload A]{\includegraphics[width
=1.7in]{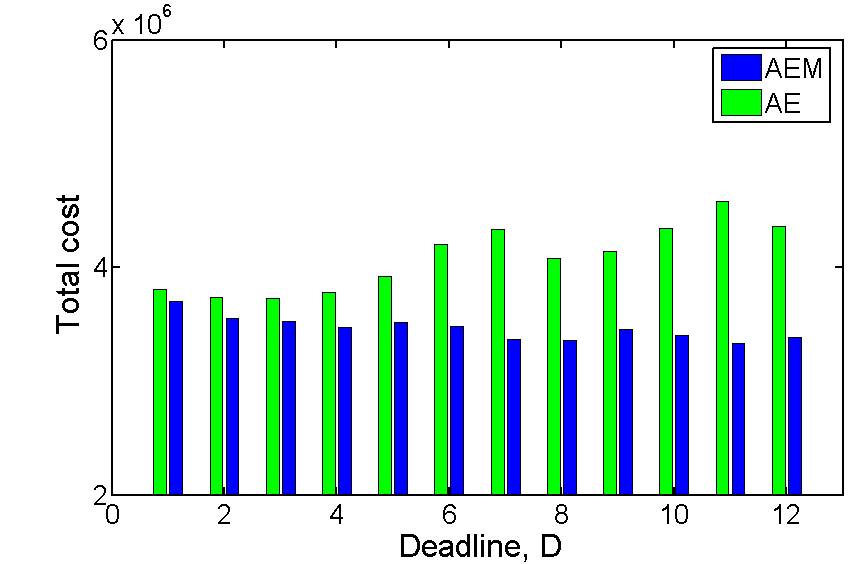}
}
\hfil
\subfigure[Workload B]{\includegraphics[width=1.7in]{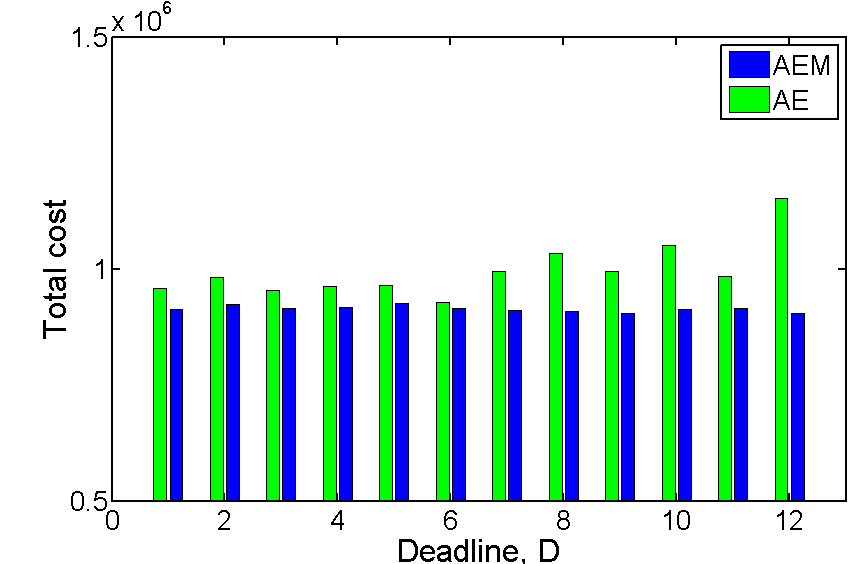}
}}
\caption{Comparison of cost incurred by the online algorithms $A_\epsilon^m$, $A_\epsilon$ (AEM and AE in figure) for different deadlines.}
\label{fig:cost_compare}
\end{figure}

We compare the cost reduction for the offline and the online algorithm $A_\epsilon^m$ with the greedy method without dynamic deferral. Figure~\ref{fig:cost_reduction} depicts the cost reduction for the online and offline algorithms for different deadlines. These curves show that dynamic deferral can provide around 30\% cost savings for deadlines of 12 slots (1 hour) and even for one slot we can get $\sim$5\% cost savings. Figure~\ref{fig:cost_compare} illustrates the comparison of the total cost from algorithms $A_\epsilon$ and $A_\epsilon^m$ for different deadlines. From this figure, we can see that the total cost from the algorithm $A_\epsilon^m$ is always less than the total cost from the algorithm $A_\epsilon$, as claimed in Lemma~\ref{lemma:online22}. As the deadline increases the total cost from the algorithm $A_\epsilon$ increases since the prediction error becomes significant for predicting more distant values (electricity prices) while the total cost from $A_\epsilon^m$ is reduced due to migration and the flexibility of dynamic deferral.

\subsubsection*{Nonuniform Deadline}
We evaluate the cost savings for nonuniform deadline assigning different deadline by classifying the workload as shown in Table~\ref{table:nonuniform}. For conservative estimates of deadline requirements (1-10), we found 15.64\% cost reduction for Workload A and 9.23\% cost reduction for Workload B each of which remains close to the offline optimal solutions.

\section{Related Work}
Greening data centers is becoming an increasingly important topic in operating cloud-scale data centers for two main reasons: (1) the global energy crisis and environmental concerns (e.g. global warming) \cite{GLB_renewable} and (2) increasing energy consumption in data centers \cite{EPA_Report}. We now discuss the related work.

{\it Energy Management in data centers.} With the importance of energy management in data centers, many scholars have applied energy-aware scheduling because of its low cost and practical applicability. Beloglazov et al. \cite{beloglazov_survey} give the taxonomy and survey on energy management in data centers. Lin et al. \cite{Dynamic_RS} have tried to minimize the energy cost together with delay cost by rightly sizing data centers. Unlike their work, we focus on distributing requests among data centers in different locations considering fixed capacity for data centers.

{\it Geographical Load Balancing.} The research community has recently identified the potential of reducing the operating cost in data centers by geographical load balancing based on the spatio-temporal variation in electricity prices. Qureshi et al. \cite{Qureshi} studied the problem of reducing the electricity cost in a wholesale market environment. They try to lower the electricity bill by utilizing the varying electricity prices in different locations of distributed data centers. They describe greedy heuristics and evaluate them on historical electricity prices and network traffic data. But they did not consider migration of workload and SLA requirements. In this paper, we utilize SLA information for load balancing and compare our algorithms with their proposed greedy algorithm. Rao et al. \cite{rao_market} consider load-balancing of delay sensitive applications with the objective of minimizing the current energy cost subject to delay constraints in a multi-electricity-market environment. They used sophisticated queuing theory to restrict the average latency rather than utilizing the flexibilities from the SLAs and dynamic migration of workload. Buchbinder et al. \cite{Buchbinder} presented online algorithms for migrating jobs between data centers, which handle the fundamental tradeoff between energy and bandwidth costs. For constant workload they could give bounded competitive ratio but for varying workload they presented a heuristic algorithm to reduce the computational complexity without making any probabilistic assumption about the future workload and future electricity prices. In contrast, we use the deadline requirements and use probabilistic assumptions to make scheduling decisions. There has also been some work on utilizing renewable energy for energy efficiency in data centers. Liu et al. \cite{GLB_renewable} presented formulation for geographical load balancing without deadline and investigated how renewable energy can be used to lower the electricity price of brown energy. In contrast we consider migration of workload between data centers to utilize energy price variation via dynamic deferral.
Le et al. \cite{Le} propose to cap the consumption of brown energy while maintaining service level agreements (SLAs). Unlike their method for satisfying SLAs, we utilize the flexibility from the SLAs for reducing energy consumption.
Stewart et al. \cite{stewart_renewable} try to maximize the use of renewable energy in data centers. However, they assume that data centers have their own energy sources (solar plants, wind mills, etc.). Unlike using renewable energy, we consider a different case where the cloud service providers buy energy from whole sale markets, which is a more common case for many data centers.

{\it Scheduling with deadline.} Many applications in real world require delay bound or deadline constraint e.g. see Lee et al. \cite{24}.  When combining with energy conservation, deadline is usually a critical adjusting tool between performance loss and energy consumption.  Energy efficient deadline scheduling was first studied by Yao et al. \cite{10}. They proposed algorithms, which aim to minimize energy consumption for independent jobs with deadline constraints on a single variable-speed processor. After that, a series of work was done to consider online deadline scheduling in different scenarios, such as discrete-voltage processor, tree-structured tasks, processor with sleep state and overloaded system \cite{11,12}. In the context of data center, most work on energy management merely talk about minimizing the average delay but not give any bound on delay except Mukherjee et al. \cite{17}. They proposed online algorithms considering deadline constraints to minimize the computation, cooling and migration energy for machines. However, their work is for job assignment inside one data center without electricity price variation.

{\it Models for electricity price prediction.} In \cite{gonzalez_hiddenmarkov}, Gonz\'{a}lez et al. presented the taxonomy of electricity price prediction models. Accordingly electricity price in wholesale market is not easy to predict due to the uncertainty of exogenous variables (e.g. energy demand, water inflow, availability of generation unit, fuel costs). People have tried predicting electricity prices using the autoregressive integrated moving average (ARIMA) model \cite{ARIMA}, generalized autoregressive conditional heteroskedasticity (GARCH) model \cite{GARCH}, wavelet transform \cite{wavelet}, dynamic regression and transform function model \cite{transfer_function}. Artificial intelligent methods that are also suitable for price forecast include artificial neural networks (ANN) \cite{ANN}, support vector machines (SVM) \cite{SVM} and Input-Output Hidden Markov Model (IOHMM) \cite{gonzalez_hiddenmarkov}. But all of these models are based on time series which are useful for predicting single value. In our algorithms we need to predict future $D$ values where the time series methods do not perform well. Hence we use Gaussian distributions to generate future values where the mean is predicted by time series method and variance is estimated from previous history. Mohsenian and Garcia \cite{mohsenian_residential} recently proposed a simple and efficient weighted average price prediction filter to predict electricity prices based on the prices from previous day and the same day in previous week. In this paper, we use this model to estimate variance because of its low computational complexity.

\section{Conclusion}
In this paper we have proposed online algorithms for geographical load balancing in data centers while guaranteeing the deadlines. The algorithms utilize the latency requirements of workloads as well as exploit the electricity price variation for cost savings and guarantee bounded cost and bounded latency under very general settings - arbitrary workload, general deadline and general energy cost models. Further the online algorithms are simple to implement and do not require significant computational overhead. To the best of our knowledge, this is the first formulation for load balancing with deadline utilizing the slackness in the execution of jobs for energy savings.

Our experiments highlight that significant cost and energy savings can be achieved via dynamic deferral of workload. However the performance of the online algorithms depend on the price variation and the quality of prediction. In this paper, we tried to limit our motivation towards the cloud considering data centers as computation units. Other factors such as capacity provisioning, heterogeneity, availability of renewable energy etc. could be taken into account during load balancing decisions. We would like to consider these issues with load balancing in future. Also it would be interesting to carry out probabilistic analysis for cost saving in demand-response market.


\begin{thebibliography}{1}
\bibitem{EPA_Report}
N. Anderson, \emph{Epa: Power usage in data centers could double by 2011}, http://arstechnica.com/old/content/2007/08/
epa-power-usage-in-data-centers-could-double-by-2011.ars, August 2007.
\bibitem{beloglazov_survey}
A.~Beloglazov, R.~Buyya, Y.~C. Lee, and A.~Zomaya, \emph{A taxonomy and survey of energy-efficient data centers and cloud computing systems}, In Advances in Computers, Elsevier: Amsterdam, 2011.
\bibitem{gandhi}
A. Gandhi, M. Harchol-Balter, R. Das, and C. Lefurgy, \emph{Optimal power allocation in server farms}, In Proc. ACM
SIGMETRICS, 2009.
\bibitem{GLB_Wierman}
Z. Liu, M. Lin, A. Wierman, S. Low, and L. H. Andrew, \emph{Greening Geographical Load Balancing,} In Proc.
ACM SIGMETRICS, June 2011.
\bibitem{Qureshi}
A. Qureshi, R. Weber, H. Balakrishnan, J. Guttag, and B. Maggs, \emph{Cutting the electric bill for internet-scale systems,} In Proc. ACM SIGCOMM, August 2009.
\bibitem{rao_market}
L. Rao, X. Liu, L. Xie, and W. Liu. \emph{Minimizing electricity cost: Optimization of distributed internet data centers in a multi-electricity-market environment}, In Proc. IEEE INFOCOM, 2010.
\bibitem{Buchbinder}
N. Buchbinder, N. Jain, and I. Menache, \emph{Online job-migration for reducing the electricity bill in the cloud,} In Proc. IFIP Networking, 2011.
\bibitem{Deadline_Ballani}
C. Wilson, H. Ballani, T. Karagiannis, and A. Rowstron, \emph{Better Never than Late: Meeting Deadlines in Datacenter Networks}, In Proc. of ACM SIGCOMM, August 2011.
\bibitem{garg_sla}
S. K. Garg and R. Buyya. \emph{SLA-based Resource Provisioning for Heterogeneous Workloads in a Virtualized Cloud Datacenter}, In Proc. of ICA3PP, October 2011.
\bibitem{stewart_renewable}
C. Stewart and K. Shen, \emph{Some Joules Are More Precious Than Others: Managing Renewable Energy in the
Datacenter,} In Proc. Power Aware Comput. and Sys., October 2009.
\bibitem{GLB_renewable}
Z. Liu, M. Lin, A. Wierman, S. Low, and L. H. Andrew, \emph{Geographical load balancing with renewables}, In Proc. GreenMetrics, June 2011.
\bibitem{Dynamic_RS}
M. Lin, A. Wierman, L. H. Andrew, and E. Thereska, \emph{Dynamic right-sizing for power-proportional data centers}, In Proc. IEEE INFOCOM, April 2011.
\bibitem{zhang2}
Y. Zhang, Y. Wang, and X. Wang., \emph{GreenWare: Greening Cloud-Scale Data Centers
to Maximize the Use of Renewable Energy}, In Proc. Middleware, 2011.
\bibitem{Le}
K. Le, O. Bilgir, R. Bianchini, M. Martonosi, and T. D. Nguyen, \emph{Managing the cost,
energy consumption, and carbon footprint of internet services}, In Proc. ACM SIGMETRICS,
2010.
\bibitem{pedram}
E. Pakbaznia and M. Pedram, \emph{Minimizing data center cooling and server power costs}, In Proc. ISLPED, 2009.
\bibitem{mohsenian_residential}
A. H. Mohsenian-Rad and A. Leon-Garcia, \emph{Optimal residential load
control with price prediction in real-time electricity pricing environments,} IEEE Trans. Smart Grid, 1(2), pp. 120-133, Sep. 2010.
\bibitem{ne_iso}
http://www.iso-ne.com.
\bibitem{nyiso}
http://www.nyiso.com.
\bibitem{ercot}
http://www.ercot.com.
\bibitem{nz}
http://www.electricityinfo.co.nz.
\bibitem{n4}
Y. Chen, A. Ganapathi, R.Griffith, and R. Katz, \emph{The Case for Evaluating MapReduce Performance Using Workload Suites,}
in Proc. IEEE MASCOTS, 2011.
\bibitem{n9}
A. Verma, L. Cherkasova, R. Campbell, \emph{SLO- Driven Right-Sizing and Resource Provisioning of MapReduce Jobs}, in Proc. LADIS, 2011.
\bibitem{n10}
A. Verma, L. Cherkasova, R. Campbell, \emph{Resource Provisioning Framework for MapReduce Jobs with Performance Goals,} in Proc. Middleware, 2011.
\bibitem{n11}
K. Kc and K. Anyanwu, \emph{Scheduling Hadoop Jobs to Meet Deadlines,} in Proc. IEEE CloudCom, 2010.
\bibitem{24}
C. B. Lee, A. Snavely, \emph{Precise and realistic utility functions for user-centric performance analysis of schedulers}, In Proc. HPDC, 2007.
\bibitem{10}
F. Yao, A. Demers, and S. Shenker, \emph{A scheduling model for reduced CPU energy}, In Proc. FOCS, pp. 374-382, 1995.
\bibitem{11}
H. L. Chan, J. W. T. Chan, T. W. Lam, L. K. Lee, K. S. Mak and P. W. Wong, \emph{Optimizing throughput and energy in online deadline scheduling}, ACM Trans. Algorithms 6(1), 1-10, 2009.
\bibitem{12}
X. Han, T. W. Lam, L. K. Lee, I. K. To and P. W. Wong, \emph{Deadline scheduling and power management for speed bounded processors}, Theor. Comput. Sci. 411(40-42), 3587-3600, 2010.
\bibitem{17}
T. Mukherjee, A. Banerjee, G. Varsamopoulos, and S. K. S. Gupta, \emph{Spatio-Temporal Thermal-Aware Job Scheduling to Minimize Energy Consumption in Virtualized Heterogeneous Data Centers}, Computer Networks, 2009.
\bibitem{gonzalez_hiddenmarkov}
A. M. González, A. M. San Roque, and J. García-González, \emph{Modeling
and forecasting electricity prices with input/output hidden markov
models}, IEEE Trans. Power Syst., 20(1), pp. 13-24, Feb. 2005.
\bibitem{ARIMA}
J. Contreras, R. Espinola, F. J. Nogales, A. J. Conejo, \emph{ARIMA
models to predict next-day electricity prices,} IEEE Trans. Power
Syst., 18(3), pp. 1014-1020, 2003.
\bibitem{GARCH}
R. C. Garcia, J. Contreras, M. van Akkeren, and J. B. C. Garcia, \emph{A
GARCH forecasting model to predict day-ahead electricity prices},
IEEE Trans. Power Syst., 20(2), pp. 867-874, 2005.
\bibitem{wavelet}
A. J. Conejo, M. A. Plazas, R. Espinola, and A. B. Molina, \emph{Day-ahead
electricity price forecasting using the wavelet transform and ARIMA
models,} IEEE Trans. Power Syst., 20(2), pp. 1035-1042, 2005.
\bibitem{transfer_function}
F. J. Nogales, J. Contreras, A. J. Conejo, and R. Espinola,
\emph{Forecasting next-day electricity prices by time series models,}
IEEE Trans. Power Syst., 17(2), pp. 342-348, 2002.
\bibitem{ANN}
B. R. Szkuta, L. A. Sanabria, and T. S. Dillon, \emph{Electricity price
short-term forecasting using artificial neural networks}, IEEE Trans.
Power Syst., 14(3), pp. 851-857, 1999.
\bibitem{SVM}
D. C. Sansom, T. Downs, and T. K. Saha, \emph{Evaluation of support
vector machine based forecasting tool in electricity price forecasting
for Australian National Electricity Market participants,} J. Electr.
Electron. Eng. Aust., 22(3), pp. 227-234, 2002.

\end{thebibliography}
\end{document}